# Three-dimensional imaging of single nanotube molecule endocytosis on plasmonic substrates

Guosong Hong, Justin Z. Wu, Joshua T. Robinson, Hailiang Wang, Bo Zhang and Hongjie Dai*

Department of Chemistry, Stanford University, Stanford, California 94305, USA

* E-mail: hdai@stanford.edu

## Abstract

**Investigating the cellular internalization pathways of single molecules or single nano-objects is important to understanding cell-matter interactions and to applications in drug delivery and discovery. Imaging and tracking the motion of single molecules on cell plasma membrane require high spatial resolution in three dimensions (3D). Fluorescence imaging along the axial dimension with nanometer resolution has been highly challenging but critical to revealing displacements in trans-membrane events. Here, utilizing a plasmonic ruler based on the sensitive distance dependence of near-infrared fluorescence enhancement (NIR-FE) of carbon nanotubes on a gold plasmonic substrate, we probe ~10 nm scale trans-membrane displacements through changes in nanotube fluorescence intensity, enabling observations of single nanotube endocytosis in 3D. Cellular uptake and trans-membrane displacements show clear dependences to temperature and clathrin assembly on cell membrane, suggesting that the cellular entry mechanism for a nanotube molecule is via clathrin-dependent endocytosis through the formation of clathrin-coated pits on cell membrane.**

## Introduction

The interactions of molecules and nanostructured materials with mammalian cells have aroused a great deal of scientific interest with implications to many biological and medical applications in drug discovery, nanomedicine and toxicology.[1] The uptake pathway and subsequent intracellular trafficking have been intensely studied and debated for a broad range of nanomaterials, including fullerene,[2] quantum dots (QDs),[3] magnetic nanoparticles[4] and carbon nanotubes (CNTs).[5-7] An interesting question is in regards to the cellular internalization pathways and the dependence of the pathways on the size of molecules or nanomaterials. It is important to investigate the upper size limit for molecules simply inserting and diffusing through the cell plasma membrane, and the lower size limit for nanomaterials becoming too small for wrapping by a highly curved lipid bilayer to undergo endocytosis.

As an example, two distinct pathways for carbon nanotube cellular entry have been suggested, direct insertion or diffusion through the lipid bilayer,[5,8-10] and clathrin-dependent endocytosis.[6,11-13] In most cases, ensembles of CNTs (including bundles and aggregates of tubes) are investigated in cellular uptake experiments through imaging of fluorescent dye labels,[5,6] and have suggested endocytotic internalizations of CNTs[6,13,14] except in several reports proposing direct insertion.[5,8] At the individual nanotube level, imaging CNT-cell interactions through detecting the intrinsic near-infrared (NIR) photoluminescence[7,15] of CNTs has been performed. However, thus far, there has been no direct three dimensional (3D) imaging of single nanotube molecules traversing cell membrane to directly observe either insertion or endocytosis pathway. In a theoretical study, Gao et al. suggested high curvature of cargo particles such as a bare single-walled CNT (diameter down to 1 nm) could cause elevated elastic energy associated with lipid-bilayer wrapping involved in endocytosis, and an optimum radius of curvature of ~14 nm exists for endosome formation around a cylindrical particle.[16] The question of whether an individual CNT (rather than bundles or aggregates of tubes) can undergo endocytosis remains an interesting open problem.

Imaging and tracking single events of specific molecules on cell membrane can offer new insights into various mechanisms of interest to biological systems.[17] Imaging single molecule trans-membrane motion requires nanometer spatial resolution along the axial dimension. Recent progresses in 3D single particle tracking have led to various new techniques to resolve the location of a single nanoparticle with high precision and elucidate interactions between the tracked nanoparticle with its surroundings.[18] For instance, by confining illumination to a ~ 100 nm optical section with evanescent waves travelling at the cover slip-cell interface, total internal reflection fluorescence microscopy (TIRFM) has been developed to image cell membrane, nearby cytoplasm and membrane-related events.[19] Further, combined with two-dimensional (2D) super-resolution techniques,[20] high axial resolution can also be achieved by imposing $z$-dependent asymmetry into two orthogonal axes $x$ and $y$,[21,22] or using two stimulated emission depletion (STED) beams to generate a central zero in three dimensions.[23] Single-molecule axial tracking has also been realized by feedback tracking, including focusing two circularly scanning laser beams at different $z$-depths.[24] Still, it remains highly challenging to image trans-membrane motion of a single molecule (such as a single nanotube) with sensitivity on the order of the thickness of the plasma membrane (~10 nm), thus probing the pathway and kinetics of single-molecule transportation across the plasma membrane.

Here we report the use of an NIR fluorescence enhancement (NIR-FE) phenomenon on plasmonic gold substrates[25-28] to track cellular internalization of individual single-walled carbon nanotubes (SWCNTs) in 3D, with an axial resolution on the order of ~ 10 nm owing to the highly sensitive dependence of fluorescence enhancement to the distance between SWCNT and gold.[25] SWCNTs exhibit intrinsic band-gap fluorescence in the 0.9-1.4 μm NIR II region upon excitation in the visible or NIR I (600-900 nm).[29-37] Recently, we observed fluorescence enhancement of SWCNTs by > 10 fold on a solution-phase grown gold films (called Au/Au films) containing patchworks of nanostructured Au islands. The fluorescence enhancement rapidly decreases as the distance between SWCNT and Au increases, with an exponential decay distance (1/$e$ decay distance) of a mere ~6 nm.[25,27] By taking advantage of this interesting effect, we demonstrate single molecule trans-membrane imaging with high sensitivity to axial motion and elucidate the cellular internalization pathway for individual nanotubes.

## Results

**Plasmonic ruler based on fluorescence enhancement.** We used water-soluble high-pressure CO conversion (HiPCO) SWCNTs (diameter ~0.7-1 nm) in our cell entry imaging experiments. The nanotubes were stably suspended by mixed surfactants of 75% C18-PMH-mPEG (5kD for each PEG chain, 90kD in total) and 25% DSPE-PEG(5kD)-NH$_2$ (see Methods) with amine groups covalently conjugated with RGD peptide ligands capable of selectively binding to α$_v$β$_3$-integrin over-expressed on U87-MG glioblastoma cells.[38] Since the radius of gyration of a 5kD PEG chain is ~3.5 nm in aqueous solution,[39] a water-soluble SWCNT is wrapped by a ~7 nm thick polymer coating radially to form a cylinder with ~15 nm diameter (inset of Figure 1b), greater than the diameter of ~1 nm of bare nanotubes. Atomic force microscope (AFM) imaging of SWCNTs on glass substrate after calcination showed most nanotubes lying horizontally with an average length of ~1 μm (Supplementary Figure S1). We found that the length distribution did not affect the major results of our experiments (Supplementary Figure S2), suggesting a general nanoscopic 'ruler' applicable to different lengths of SWCNTs.

Figure 1a shows a scanning electron microscope (SEM) image of an NIR fluorescence enhancing Au/Au film comprised of gold nano-islands with abundant gaps in between. The UV-Vis-NIR extinction spectrum of the same Au/Au film (Figure 1a inset) shows a plasmon resonance peak located at ~800 nm, facilitating fluorescence enhancement in the NIR region.[40] To reveal the distance dependence of fluorescence enhancement, we coated Au/Au films with progressively thicker $Al_2O_3$ by atomic layer deposition (ALD), and deposited nanotubes on the substrates by drop-drying an aqueous nanotube suspension. The surface density of nanotubes within the drop-dried spot can be seen from the AFM image in Supplementary Figure S1. Comparing the nanotube fluorescence intensity on Au/Au films with different spacer thicknesses to that on bare glass, we found a decreasing trend of enhancement factor, from ~8-fold enhancement to almost no enhancement, as the spacer increases in thickness (Figure 1b). The exponential fit showed a surprisingly short $1/e$ decay distance of ~6 nm considering the average length of our SWCNTs is much greater, ~ 1µm. Since the nanotubes are long and may not lie perfectly flat on the substrate, the separation of any part of a long tube from the gold surface is greater than or equal to the thickness of $Al_2O_3$ spacer. For this reason, we suggest that the measured enhancement vs. distance data corresponds to the minimum distance between Au and the length of a nanotube in the case when the nanotube is non-parallel to the Au surface.

**3D tracking of single nanotubes at 37 ℃.** We exploited the ultra-sensitive fluorescence enhancement of SWCNTs to gold-nanotube distance for probing motion of nanotubes in the direction normal to the gold surface. We stained trypsinized U87-MG cells at 4 ℃ by highly diluted SWCNTs (~20 pM) with PEG and RGD functionalization and placed the cells on an Au/Au substrate kept at 4 ℃. Imaging with an InGaAs 2D detector (excitation ~658 nm) in the 1.1-1.7 µm emission range revealed a brightly fluorescent spot overlaying with a single cell (Figure 1c inset), attributed to a single nanotube sandwiched between Au/Au and cell membrane on the proximal side to the substrate (see Supplementary Figure S3 for further evidence of bright nanotubes sandwiched between cell and Au), exhibiting strong NIR fluorescence enhancement due to proximity to the gold surface.[27] Due to trypsinization and staining at 4 ℃ to prevent unwanted endocytosis before tracking started, most cells lost extracellular matrix and turned into round shape, but they still remained viability and could internalize nanomaterials from the outside once temperature allowed.[27] We identified the single nanotube by one single peak in the emission spectrum [~1150 nm in Fig.1c, corresponding to (7,6) chirality] and the sinusoidal dependence[41] of fluorescence on the polarization of laser excitation (Figure 1d).

Once an individual nanotube was identified, we increased the incubation temperature from 4 ℃ to 37 ℃ *in situ* and tracked the fluorescence of the SWCNT over time at a frame rate of 0.3 frame/sec after the temperature stabilized at 37 ℃. The (7,6) tube in Fig.1c showed a clear fluorescence intensity decrease over time (see Figure 2b, 2c and 2d, extracted from Supplementary Movie 1) with motion in the 2D imaging plane on the cell membrane (see Figure 2a and bottom left panel of Supplementary Movie 1). By plotting the fluorescence intensity of this tube as a function of time, we observed an exponential decrease with a $1/e$ decay time of ~121 s following some fluctuations at beginning. Two other independent tracking experiments were carried out under the same condition and similar fluorescence decreases were observed (Supplementary Figure S4i-l) with $1/e$ decay times of ~103 s and ~111 s respectively.

To rule out the possibility of chemical environment change such as pH change as the main cause of the fluorescence intensity decrease, we immobilized SWCNTs with the same PEG coating on Au/Au surface without cells and adjusted the pH of the immersion medium from 5 to 9, with pH ~ 5 mimicking the acidic environment of endosomes and lysosomes.[42] A mere 15% decrease in fluorescence intensity was observed at pH ~ 5 (Supplementary Figure S5), indicating the ~7-fold decay was mostly attributed to the *z*-axis displacement while the change of chemical environment had very little contribution.

In a control experiment, we depolarized the linearly-polarized laser excitation to image and track a single nanotube endocytosis process and observed similar effect of fluorescence decrease over time with a $1/e$ decay time of 114 s at 37 ℃ (Supplementary Figure S6). This suggested the observed fluorescence decay in the linear polarization case was due to reduced fluorescence enhancement during endocytosis without large nanotube rotational effects. SWCNTs have been reported to rotate freely in water.[43] However, in our case the nanotubes showed small rotations during endocytosis due to confinement and interactions with the membrane and the long 3s integration time that had averaged out the rotational effects.

It is also possible that some nanotubes were bound to the cell membrane perpendicular to the Au-cell interface initially and endocytosed vertically as suggested recently for large multi-walled nanotubes.[44] The laser excitation used in our experiments was not able to excite and resolve these tubes efficiently. Also noticeable is that we had not observed a brightly fluorescent nanotube evolving into a dim state in a time scale of several seconds (the estimated rotation time if assuming free rotation in a viscous medium corresponding to the slow translational diffusivity measured), suggesting none of the SWNTs imaged had changed orientation from in the *x-y* plane to pointing to the z-direction during endocytosis.

Given the control experiments ruling out other possibilities of fluorescence decay, the observed fluorescence decrease of single nanotubes on cells at 37 ℃ hinted axial motion of nanotubes away from the Au/Au substrate due to ultra-sensitive Au-SWCNT distance dependence of nanotube fluorescence intensity. Based on the nanoscopic ruler effect of fluorescence enhancement, we rationalized that during 4 ℃ staining, RGD functionalized SWCNTs selectively attached to the $\alpha_v\beta_3$-integrin receptors on the cell membrane (Figure 2f) without entering the cytoplasm due to blocked endocytosis at 4 ℃.[6] At 37 ℃, endocytosis was activated with the formation of a vesicle wrapping around the surfactant-coated nanotube via clathrin-associated invagination of the plasma membrane (Figure 2g), followed by vesicle pinching-off (Figure 2h) and clathrin uncoating to undergo the endocytotic pathway.[45]

In the first ~ 20 s at 37 ℃, the nanotube was sandwiched between the plasma membrane and the underlying Au/Au film (Figure 2f), exhibiting high fluorescence due to close proximity to the fluorescence enhancing surface (Figure 2b). The distance from the membrane to substrate could reach 4~8 nm[46] in less than 5 min[47] according to the Derjaguin–Landau–Verwey–Overbeek (DLVO) model.[48] Over time (from Figure 2b to 2c), clathrin assembled on the inner surface of plasma membrane and the membrane bent inwards to form clathrin-coated pits and wrapped around the nanotube. Due to invagination, the nanotube was displaced away from the Au substrate (Figure 2g), leading to reduced fluorescence enhancement (Figure 2c). At later times (from Figure 2c to 2d), the clathrin-coated pit continued to grow and finally pinched off to form a complete vesicle enclosing the nanotube. At this point the nanotube was >20 nm away from the Au substrate due to the two lipid bilayers in between (Figure 2h). This large spatial separation led to very little enhancement from the plasmonic substrate (Figure 2d). A complete sequence of such regulated events including clathrin assembly, pits formation, budding and clathrin uncoating usually takes tens of seconds to a few minutes at 37 ℃, depending on the size of the cargo molecule.[45] In the case of this particular SWCNT, the time required to complete fluorescence decrease was ~ 250 s (Figure 2e), within the reported time range to complete clathrin cluster assembly[49] but on the higher side, similar to the time needed to internalize relatively large cargo molecules (~400 s) such as reovirus particles (~85 nm in diameter).[45]

**3D tracking of single nanotubes at various temperatures.** To support that trans-membrane motion of single nanotubes at 37 °C was indeed the cause of fluorescence decrease, we imaged single nanotubes on U87-MG cells at several other temperatures, including 4 °C, 25 °C and 42 °C. It is known that at temperatures lower than 37 °C, cell functions such as active uptake are impaired, and endocytosis is completely blocked at 4 °C.[6] On the other hand, cell functions are more active at elevated temperatures until excessive heating begins to cause damages.[50] In a typical experiment at 4 °C, we observed a single nanotube (evidenced by polarization dependence in Figure 3a) moving on cell membrane (2D trajectory shown in Figure 3b). This nanotube exhibited fluctuations in fluorescence intensity without any significant decay over sufficiently long periods of time (>900 s, see Figure 3c and top left panel of Supplementary Movie 1). This was consistent with blocked endocytosis at this low temperature. The random fluctuations of fluorescence could be due to the stochastic fluctuations in the electrostatic environment of the cell medium (see Supplementary Figure S7 for the degree of fluctuation even in the absence of cell),[51] as well as small wiggles of SWCNT or subtle membrane motions that caused small Au-SWCNT distance fluctuations.

At 25 °C, a temperature sufficiently high to initiate active cellular uptake but lower than 37 °C, we observed much longer time (1/$e$ decay time = 327 s) than at 37 °C (1/$e$ decay time = 100~120 s) to complete trans-membrane displacement of nanotubes (Figure 3d-3f and top right panel of Supplementary Movie 1). At 42 °C, a temperature known to enhance the endocytosis of mammalian cells,[52] the nanotube fluorescence intensity decreased so precipitously that the bright spot disappeared into the cells within only ~50 s (Figure 3g-3i and bottom right panel of Supplementary Movie 1). Once in the cells (Supplementary Figure S8a-e) with little fluorescence enhancement by the Au/Au substrate, the nanotube could be imaged after focusing the laser beam down to a much smaller area and thus increasing the laser excitation power density by 20 times (Supplementary Figure S8f).

Importantly, control experiments were performed at various temperatures (4-42 °C) on bare glass without any gold coating (Supplementary Figure S9 and Supplementary Movie 2) for cells stained with SWCNTs. No fluorescence decrease of individual nanotubes was observed as on Au/Au plasmonic substrates at all temperatures up to long imaging and tracking times. This result again suggested that on the Au/Au plasmonic substrate, fluorescence intensity decrease of single tubes was due to reduced fluorescence enhancement as nanotubes were endocytosed into the cells away from the plasmonic substrate.

**Single nanotube endocytosis activation barrier.** The imaging experiments on plasmonic Au/Au substrates at various temperatures were repeated with different single tubes (Supplementary Figure S4) to obtain averaged $1/e$ decay times of individual nanotube fluorescence on cells (Figure 4a). Since endocytosis is a well-established energy-dependent process, from the $1/e$ decay times at different temperatures we extracted an activation barrier for the internalization process to be 120±37 kJ/mol (Figure 4b), which was significantly higher than the reported values for clathrin-mediated endocytosis of some proteins [~55 kJ/mol for mannosylated albumin (~67 kDa) in trout hepatocytes,[53] and 40±13 kJ/mol for horseradish peroxidase (~44 kDa) in salmon head kidney cells[54]]. The higher activation energy for cellular internalization of a single nanotube could be due to the large molecular weight (~1 MDa) or long tube length. To gain a quantitative view, we calculated the effective capture radius as follows,[15]

$$R^* = \frac{a}{\ln\dfrac{2a}{b}} = \frac{500 \text{ nm}}{\ln\dfrac{2 \times 500 \text{ nm}}{7.5 \text{ nm}}} = 102 \text{ nm}$$

where the average length of our SWCNTs was $2a = 1$ μm, and they were coated with long PEG chains with an overall radius of $b$ ~7.5 nm. The as-calculated effective radius lied in the vesicle wrapping region for nanoparticles but was on the high side, as suggested by the model of Gao et al.[16] An optimum size for the cellular entry of both Au nanoparticles and SWCNTs has been reported[15,55] to be ~25 nm in effective radius, smaller than the effective radius of the SWCNTs used in the current work. Therefore, although endocytosis mediated by membrane wrapping was more favorable than direct insertion in our case, the large effective capture radius slowed down this process with a relatively high apparent activation barrier.

**Mean-square-displacement (MSD) analysis of 2D trajectories.** To verify the fluorescence intensity decrease in each case indeed reflected the process of endocytosis, we also carried out the MSD analysis based on the 2D nanotube trajectories in the *x-y* plane. MSD was plotted in Figure 4c as a function of lag time for each of the four single nanotube trajectories at different temperatures in Figure 2 and 3. For nanotube trajectories recorded at three elevated temperatures 25 ℃, 37 ℃ and 42 ℃, the MSD curves all exhibited nearly linear increase initially with a small quadratic component regressed as the dashed curves. This indicated dominant Brownian motion during the vesicle formation process.[56] Diffusivity values extracted from the curves were in a range between $10^3$ to $10^4$ nm$^2$/s, consistent with the diffusion constant of membrane receptors on the order of $10^4$ nm$^2$/s [57] and suggesting nanotubes bound to membrane receptors during the initial stage of motion. The measured translational diffusivity range was 2-3 orders of magnitude smaller than in water (>1 μm$^2$/s)[43], suggesting strong interactions between nanotubes and cell membranes that hindered the diffusion or rotation of SWNT during endocytosis.

Interestingly, after sufficiently long period of time depending on the temperature, a turning point showed up in each of these three cases (940 s at 25 ℃, 260 s at 37 ℃ and 50 s at 42 ℃), after which a dramatic increase in MSD was observed. The MSD was fit to a quadratic function with reduced Brownian component, indicating the completion of vesicle formation and an active transport of the encapsulated cargo.[7] In contrast, when tracking a single nanotube at 4 ℃, the MSD curve kept increasing almost linearly without an abrupt upward turning point (Figure 4c). These results are in agreement with other studies,[7] where convective diffusion is usually observed after a single nanoparticle has been fully endocytosed due to microtubule-dependent transport by motor proteins.[58] Figure 4d revealed a large increase in convective motion velocity immediately following endocytosis. Moreover, since the turning point indicated when the complete vesicle was formed and motor proteins started transporting the cargo vesicle, we compared the turning point in each MSD curve with the time required for the starting fluorescence intensity to drop to baseline level from Figure 2 and 3, which acted as another measure of the endocytosis completion time. Strikingly, results from these two methods matched very well. At 25 ℃, the turning point was at 940 s, while endocytosis completion was at ~880 s according to fluorescence intensity decrease. At 37 ℃, the turning point was at 260 s, while the fluorescence decay curve showed complete endocytosis at ~250 s. At 42 ℃, the time of turning point was 50 s, while from the fluorescence decay the time of complete endocytosis was ~60 s. MSD analyses for all other trajectories of SWCNTs were also carried out and shown in Supplementary Figure S10. We reproducibly observed two distinct steps for the imaged SWCNTs, an initial stage of dominant Brownian motion, and subsequent convective diffusion with abrupt increase in MSD. Thus, the MSD analysis confirmed the occurrence of endocytosis, and the turning points from the MSD curves further verified our intensity-based measurement as a way to follow the endocytosis process.

**Block of endocytosis.** Previous work has found that cellular internalization of ensembles of CNTs involves the clathrin-dependent endocytosis pathway.[6] Potassium depletion, as well as hypertonic medium incubation, are the two methods known to perturb endocytosis by removing membrane-associated clathrin lattice.[59] To glean if trans-membrane endocytosis of individual SWCNTs observed with the plasmonic 'ruler' did involve the formation of clathrin lattice on the plasma membrane, we investigated the effects of potassium depletion and hypertonic treatment to the internalization of single nanotube molecules. In a hypertonic treatment, we pre-incubated U87-MG cells with sucrose-supplemented buffer at 37 ℃ and then stained the cells at 4 ℃. After the temperature of imaging chamber was raised to 37 ℃, a bright individual nanotube was found to move around (Figure 5a and left panel of Supplementary Movie 3) in 2D without fluorescence decrease over long period of time (>1000 s, Figure 5b). This result suggested blocking of endocytosis of single nanotubes by hypertonic treatment. Similar result was observed after treating cells in potassium depleted medium (Figure 5c, 5d and right panel of Supplementary Movie 3), also confirming blocking of endocytosis by depleting $K^+$ ions from the medium. It has been shown that both hypertonic incubation and potassium depletion induce abnormal clathrin polymerization into empty microcages[59] (Figure 5e), preventing interactions between clathrin and its adaptor AP-2 and thus blocking the formation of clathrin-coated pits that are essential to receptor-mediated endocytosis.

## Discussion

This work exploited the sensitive distance dependence of NIR fluorescence enhancement of single carbon nanotube molecules on a gold plasmonic substrate to probe ~10 nm scale trans-membrane displacements through changes in nanotube fluorescence intensity, presented the first 3D tracking of individual SWCNTs and established the nanotube entry pathway to be clathrin-dependent endocytosis. Compared to other existing 3D single particle imaging and tracking techniques that either require sophisticated implementation or are limited with insufficient axial and/or temporal resolution,[18] the sensitive distance dependence of fluorescence enhancement based on the plasmonic effect presents a facile, inexpensive and sensitive probing of sub-10 nm distance changes in biological systems. Notably, spatial sensitivity and range of distance measurements are difficult to optimize simultaneously. Our current method allows for probing subtle distance changes of <10 nm along the z-axis, but tracking displacements beyond ~20 nm becomes difficult due to the short plasmonic ruler length. It is necessary to resort to other fluorophores with different plasmonic ruler lengths, such as synthetic dyes, fluorescent nanoparticles or fluorescent proteins to extend the measurable range along the *z*-axis. Our preliminary results have indeed identified that the plasmonic fluorescence enhancement vs. distance profile is characteristic to each fluorophore and also depends on the type of plasmonic substrates. Suitable combinations of fluorophore-substrate could lead to a library of 'nanoscopic rulers' spanning various ranges of distances to probe molecular motions at the nanoscale.

We envisage that the reverse process of endocytosis, exocytosis[7,60] of single nanotubes or molecules could also be investigated by utilizing fluorescence enhancement phenomena on plasmonic substrates. Further, similar to trans-membrane processes, imaging with suitable plasmonic rulers may also offer a platform to decipher important biological pathways involving translocation motions of molecules inside cells. Distance information and protein conformational changes inside the cytoplasm could also be revealed by introducing plasmonic metal nanoparticles and fluorophores inside live cells.

## Methods

**Preparation of Au/Au films.** The solution-phase Au/Au film synthesis can be found in detail in another publication of our group.[26] Briefly, a glass slide was immersed in a 25 mL solution of 3 mM chloroauric acid, to which 400 μL of ammonia was added under vigorous agitation. The substrate was then allowed to sit in the seeding solution with gentle shaking for 1 min, after which it was rinsed with water. Then the substrate was submerged into a 25 mL solution of 1 mM sodium borohydride on an orbital shaker at 100 rpm for 5 min. Following a second rinsing step, the seeded substrate was soaked in a 25 mL solution of 1 mM chloroauric acid and 1 mM hydroxylamine under agitation for 15 minutes. It was then rinsed with water and soaked in 1 mM cysteamine ethanol solution for 1 h to render it hydrophilic and biocompatible.
**UV-Vis-NIR absorbance measurements.** UV-Vis-NIR absorbance curve of the as-made Au/Au film on glass substrate was measured by a Cary 6000i UV-Vis-NIR spectrophotometer, background-corrected for any glass contribution. The measured range was 400-1200 nm.
**Scanning electron microscopy (SEM) imaging.** Au/Au film grown on glass was imaged via SEM. Image was acquired on an FEI XL30 Sirion SEM with FEG source at 5 kV acceleration voltage.

**Atomic layer deposition (ALD) process.** Low temperature ALD process was used to coat the as-made Au/Au film with desirable thicknesses. Deposition was carried out on amine group functionalized hydrophilic Au/Au substrate at 100 ℃ in ~300 mTorr pure nitrogen environment where trimethylaluminum (TMA) and water vapor were used as precursors. For each cycle of ALD, water pulse was 0.5 s in duration, followed by a 40 s purging time, a 0.5 s TMA pulse and a 30 s purging time. To deposit the $Al_2O_3$ layers of 5nm, 10 nm, 15 nm and 20 nm, 50, 100, 150 and 200 cycles were used, respectively.

**Preparation of water soluble SWCNT-PEG-RGD bioconjugate.** The preparation of water soluble SWCNT fluorophores can be found in detail in another publication of our group with some modification.[34] In general, raw HiPCO SWCNTs (Unidym) were suspended in 1 wt% sodium deoxycholate aqueous solution by 1 hour of bath sonication. This suspension was ultracentrifuged at 300,000 g to remove the bundles and other large aggregates. To further remove any remaining bundles and keep only bright single nanotubes, a gradient separation was used to purify the as-made SWCNTs. The supernatant was first concentrated and then layered to the top of a 10 wt%/20 wt%/30 wt%/40 wt% sucrose step gradient, followed by ultracentrifugation at 300,000 g for 1 hour. Only the top 1 mL was retained by careful fractionation and 0.75 mg/mL of C18-PMH-mPEG (5 kD for each PEG chain, 90 kD in total) (poly(maleic anhydride-*alt*-1-octadecene)-methoxy(polyethyleneglycol, 5000) MW = 90,000 in total), synthesized by our group) along with 0.25 mg/mL of DSPE-PEG(5 kD)-$NH_2$ (1,2-distearoyl-*sn*-glycero-3-phosphoethanolamine-N-[amino(polyethyleneglycol, 5000)], Laysan Bio) was added. The resulting suspension was sonicated briefly for 5 min and then dialyzed at pH 7.4 in a 3500 Da membrane (Fisher) with a minimum of six water changes and a minimum of two hours between water changes. To remove aggregates, the suspension was ultracentrifuged again for 1 hour at 300,000 g. This surfactant-exchanged SWCNT sample has lengths ranging from 100 nm up to 3.0 µm, with the average length of ~1 µm as shown in Supplementary Figure S1. These amino-functionalized SWCNTs were further conjugated with RGD peptide according to the protocol that has been used in our group. Briefly, an SWCNT solution with amine functionality at 300 nM after removal of excess surfactant, was mixed with 1 nM sulfo-SMCC at pH 7.4 for 2 h in PBS at pH 7.4. After removing excess sulfo-SMCC by filtration through 100-kDa filters (Amicon), RGD-SH (cyclo-RGDFC, Peptides International) was added together with tris(2-carboxyethyl)phosphine (TCEP) at pH 7.4. The final concentrations of SWCNT, RGD-SH and TCEP were 300 nM, 0.1 mM and 1 mM, respectively. The reaction was allowed to proceed for 2 days at 4 ℃ before purification to remove excess RGD and TCEP by filtration through 100-kDa filters.

**Atomic force microscopy (AFM) imaging.** AFM image of the as-made SWCNT conjugate was acquired with a Nanoscope IIIa multimode instrument in the tapping mode. The sample for imaging was a drop-dried sample on glass, the same one for calibration curve measurement of the distance dependence of fluorescence enhancement, prepared by drop-drying 0.5 µL water-soluble PEGylated and functionalized SWCNTs (0.45 nM) solution containing 0.05 wt% Triton X-100 on a bare glass substrate and calcination at 350 ℃ for 15 min.

**Distance-dependence of plasmonic fluorescence enhancement.** To determine the calibration curve, on the bare glass substrate and each Au/Au substrate with a certain thickness of $Al_2O_3$ coating, 0.5 μL water-soluble, PEGylated and functionalized SWCNTs (0.45 nM) solution containing 0.05 wt% Triton X-100 was drop-dried to form a uniform spot with diameter of ~2 mm. All spots were imaged in epifluorescence setup with a 658-nm laser diode (100 mW, Hitachi) focused to a 750 μm diameter spot by focusing the laser near the back focal length of a ×10 objective lens (Bausch & Lomb). The resulting NIR photoluminescence (PL) was collected using a liquid-nitrogen-cooled, $320 \times 256$ pixel, two-dimensional InGaAs camera (Princeton Instruments) with a sensitivity ranging from 800 to 1,700 nm. The excitation light was filtered out using an 1,100 nm long-pass filter (Omega) so that the intensity of each pixel represented light in the $1,100 – 1,700$ nm range. The exposure time was 100 ms. Images were taken in a 2D scanning mode, flat-field-corrected to account for non-uniform laser excitation, and then stitched automatically using LabVIEW to recover the original shape of the spot. Integral intensity for each stitched spot was done using the *roipolyarray* function in MATLAB software.

**Cell incubation and staining.** The U87-MG cell medium containing 1 g Ł$^{-1}$ D-glucose, 110 mg Ł$^{-1}$ sodium pyruvate, 10% fetal bovine serum, 100 IU mL$^{-1}$ penicillin, 100 μg·mL$^{-1}$ streptomycin and L-glutamine was used. Cells were maintained in a 37 ℃ incubator with 5% $CO_2$. To stain cells with SWCNT-PEG-RGD, trypsinized U87-MG cells were mixed with SWCNT-PEG-RGD at a nanotube concentration of 1 nM (for imaging on glass; this higher concentration helped to find the brighter single nanotubes from a distribution including many dim tubes) or 20 pM (for imaging on Au/Au which helped to visualize the otherwise dim tubes by enhancement) at 4 ℃ for 1 h, followed by washing the cells with 1x PBS to remove all free conjugates in the suspension. For the hypertonic treatment, the original cell medium was completely removed and the cells were incubated in 1x PBS supplemented with 0.45 M sucrose at 37 ℃ for 0.5 h before trypsinized and stained. For the potassium depletion treatment, the original cell medium was removed and the cells were incubated in a potassium-free buffer containing 0.1 M HEPES (4-(2-hydroxyethyl)-1-piperazineethanesulfonic acid), 1.4 M NaCl and 25 mM $CaCl_2$ at 37 ℃ for 0.5 h before trypsinized and stained. Note that for the K$^+$-depletion treatment, the potassium-free buffer was used to replace 1x PBS wherever 1x PBS was needed throughout the entire procedure.

**High-magnification NIR photoluminescence imaging.** 5 μL of stained U87-MG cell suspension was transferred to 200 μL of 1x PBS or potassium-free buffer (for potassium depletion treatment), placed into an 8-well chamber slide (Lab-Tek™ Chambered Coverglass, 1.0 Borosilicate). An Au/Au coated glass substrate or a bare glass chip was then placed on top. Capillary force formed a very thin layer of liquid between the substrate and coverglass, allowing a monolayer of cells residing in between. The chamber slide was kept in a temperature controlled chamber (BC-260W, 20/20 Technology, Inc.) for epifluorescence imaging. Temperature was always kept at 4 ℃ at the beginning for at least 5 min to ensure the cell membrane in close contact with the hydrophilic gold surface. The temperature of the imaging cell was controlled by heat exchanger (HEC-400, 20/20 Technology, Inc.), and the $CO_2$ gas flow was kept at 1 L/min by a gas purging system (GP-502, 20/20 Technology, Inc.). Single nanotube imaging and tracking were done using a 658-nm laser diode excitation with an 80 μm diameter spot focused by a ×100 objective lens (Olympus). The resulting NIR photoluminescence was collected using a liquid-nitrogen-cooled, 320 × 256 pixel, two-dimensional InGaAs camera (Princeton Instruments) with a sensitivity ranging from 800 to 1,700 nm. The excitation light was filtered out using an 1100 nm long-pass filter (Thorlabs) so that the intensity of each pixel represented light in the 1,100 − 1,700 nm range. To initiate the active uptake of single nanotubes, the chamber temperature was increased from the initial temperature of 4 ℃ and stabilized at the set temperature (usually took 2 min). The camera took images to record endocytotic process continuously with an exposure time of 3 s. Matlab 7 was used to process the images for any necessary flat-field correction and extract trajectories and time courses of PL changes from the video.

## Acknowledgements

This work was supported by NIH-NCI 5R01CA135109-02 and Stanford Graduate Fellowship.

## Author contributions

H.D. and G.H. conceived and designed the experiments. G.H., J.Z.W., J.T.R., H.W. and B.Z. performed the experiments. G.H. and H.D. analysed the data and wrote the manuscript. All authors discussed the results and commented on the manuscript.

## Figure legends

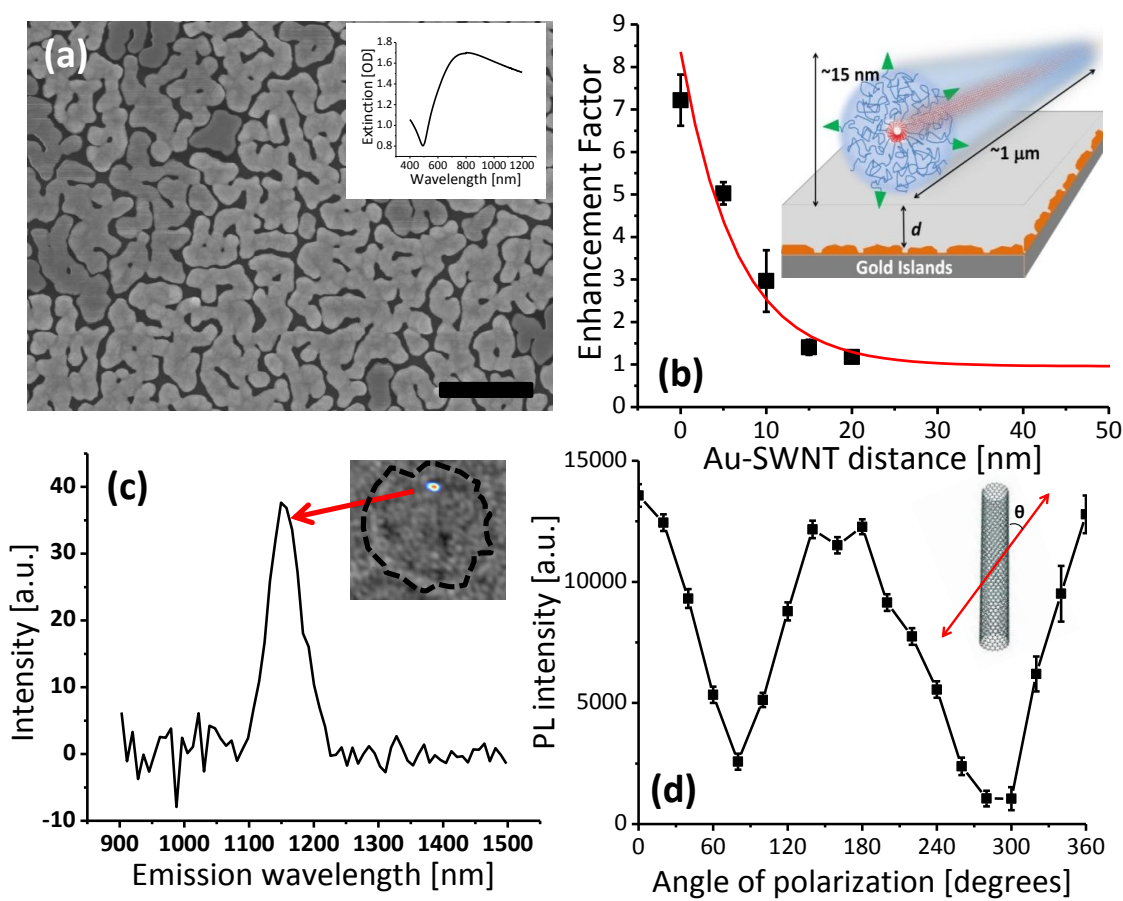

**Figure 1.** Single nanotube-cell imaging on a plasmonic gold substrate. (a) An SEM image of the plasmonic Au/Au film used for enhancing the fluorescence of SWCNTs in this work. The scale bar indicates 500 nm. Inset shows the UV-Vis-NIR extinction spectrum for the same film with a plasmonic resonance peak at ~800 nm. (b) Fluorescence enhancement factor of SWCNTs as a function of distance

between SWCNTs and the enhancing Au/Au film showing a 1/$e$ decay distance of ~6 nm. Data are shown as black squares with error bars obtained by taking the standard deviation of all 319×256 pixels in the field of view (690 µm by 860 µm), while a fit to first-order exponential decay is shown in red curve. The schematic shows how a single nanotube (red cylindrical core) wrapped with polymer coating (cyan shell) is separated from the Au/Au film by an Al$_2$O$_3$ spacing layer with a certain thickness $d$. RGD peptide ligands are shown in green triangles. (c) Emission spectrum of a fluorescence-enhanced single CNT under excitation at 658 nm, showing a peak at 1149 nm. The nanotube is assigned to the (7,6) chirality. Inset: overlay of fluorescence image and white-light U87-MG cell image (outlined by the dashed line) showing the single tube on the cell membrane at the proximal side to the Au/Au film. (d) Polarization dependence of the photoluminescence (PL) intensity of the SWCNT in (c). Error bars were obtained by taking the standard deviation of the brightest and eight surrounding pixels that make up the single nanotube image. Inset shows that the angle of polarization is defined as the angle between the electric field vector of polarized excitation light and the nanotube axis.

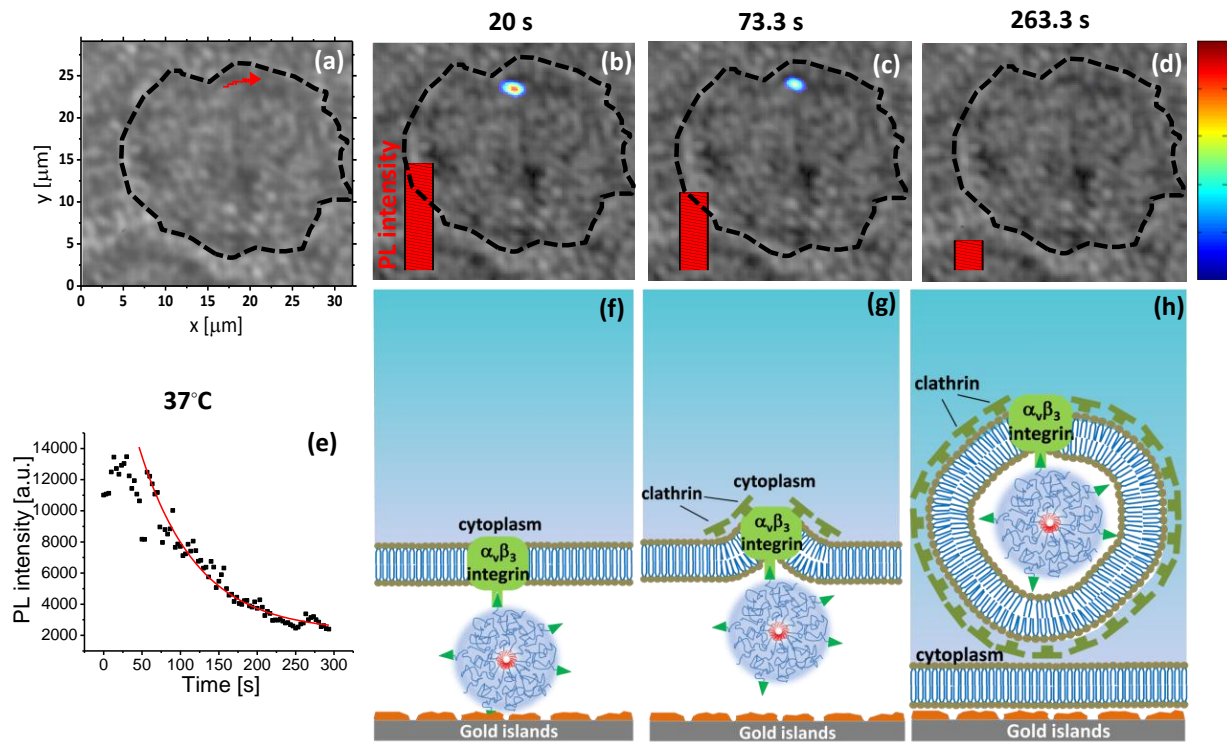

**Figure 2.** Single nanotube imaging and tracking on cells on a plasmonic gold substrate at 37 ℃. (a) 2D trajectory of a single CNT (same as the one in Figure 1c and 1d) moving on the cell membrane during endocytosis at 37 ℃. (b-d) Photoluminescence images of this particular single CNT overlaid with optical images of a cell (outlined by the dashed line) at three time points: 20 s, 73.3 s and 263.3 s after the incubation temperature stabilized at 37 ℃. Rectangular red bar on the bottom left of each image indicates relative fluorescence intensity. Intensity scale bar on the right indicates red as the highest signal while blue as the lowest after normalization. (e) Photoluminescence (PL) intensity (black squares) of this single nanotube plotted as a function of time at 37 ℃ and fitted to a first-order exponential decay shown in red curve with a 1/e decay time of 121 s. (f-h) Schematics of sequential stages during endocytosis of the individual nanotube with surfactant coating. Only cross-section of the single nanotube is shown in the drawings.

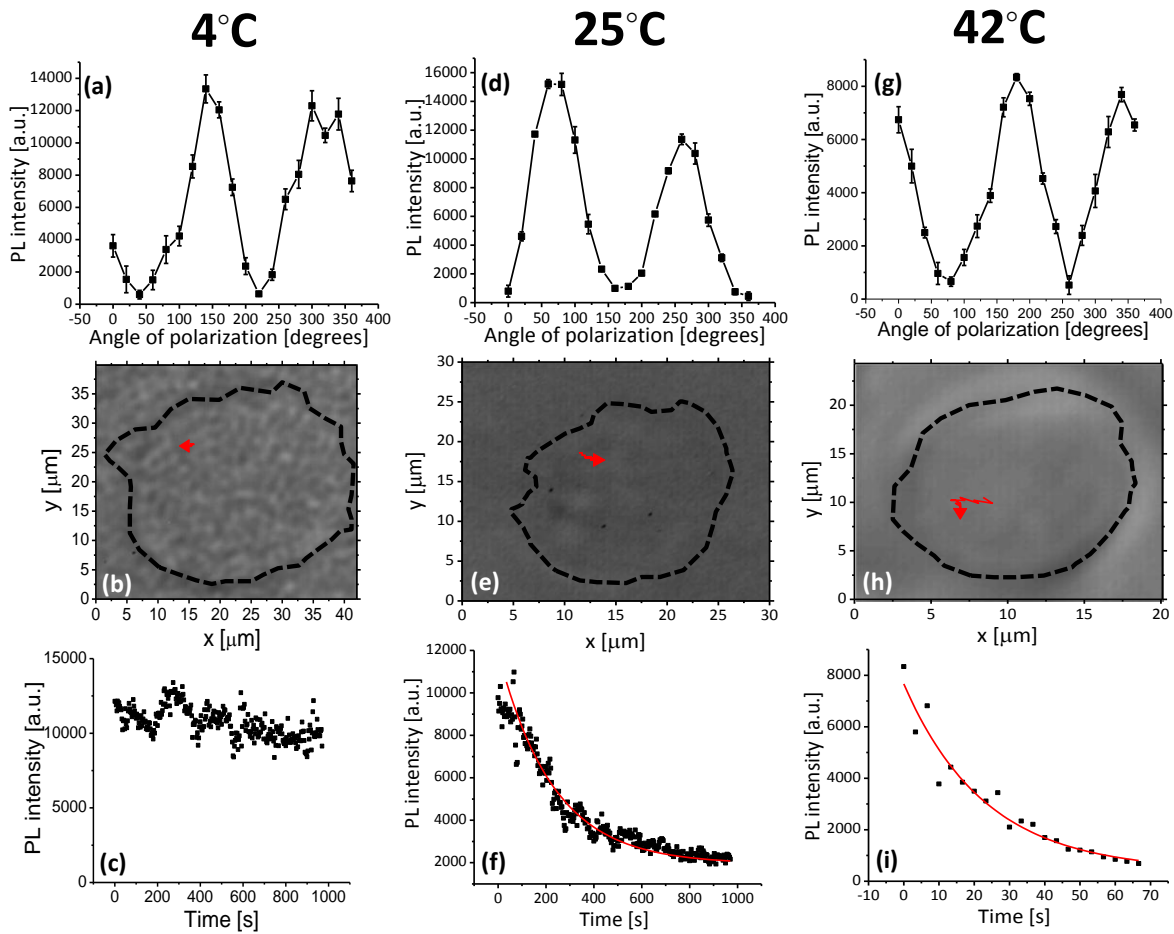

**Figure 3.** Single nanotube imaging and tracking on cells on plasmonic gold substrates at 4 ℃, 25 ℃ and 42 ℃. (a), (d) and (g): Proof of single nanotubes being imaged through polarization dependent fluorescence data. Error bars were obtained by taking the standard deviation of the brightest and eight surrounding pixels that make up the single nanotube image. (b), (e) and (h) show 2D trajectories of three different, individual SWCNTs on cell membranes at 4 ℃, 25 ℃ and 42 ℃ respectively. (c), (f) and (i) show the photoluminescence (PL) intensity (black squares) over time for the three individual nanotubes on cells at 4 ℃, 25 ℃ and 42 ℃ respectively. At 25 ℃ and 42 ℃ the data were fitted into a first-order exponential decay (red curve) with a 1/e decay time of 327 s and 23 s, respectively.

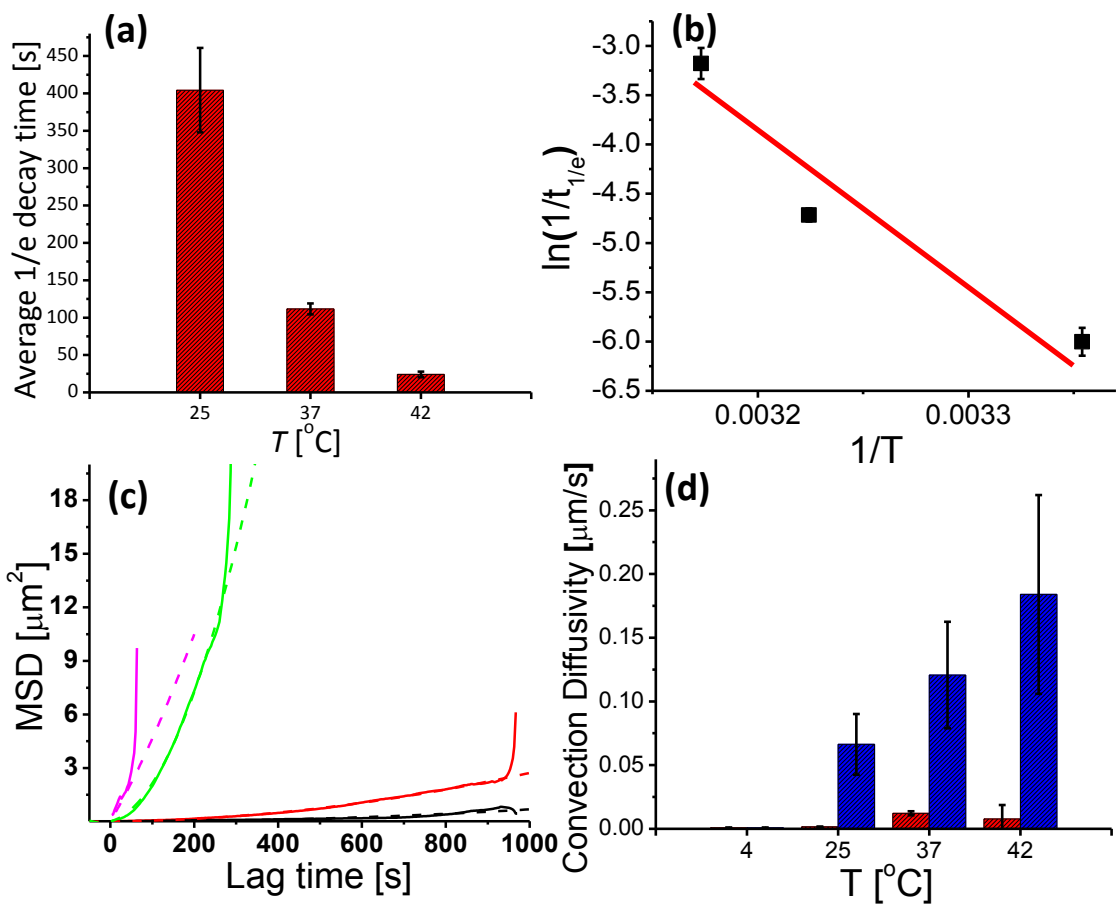

**Figure 4.** Temperature-dependent endocytosis time and in-plane diffusion. (a) A bar chart showing the average $1/e$ SWCNT fluorescence decrease (caused by trans-membrane displacements during endocytosis) times at 25 ℃, 37 ℃ and 42 ℃, corresponding to the averaged endocytosis times of SWCNTs at these temperatures. Error bars were obtained by taking the standard deviation of three independent experiments at each temperature. (b) An Arrhenius plot for determining the activation energy of single nanotube trans-membrane motion. Data points shown as black squares were fitted into a linear equation from which the activation barrier was extracted from the slope as ~120 kJ/mol, and the error bars were obtained by taking the standard deviation of three independent experiments at each temperature. (c) Mean-square-displacement (MSD) curves vs. lag time from four single nanotube tracking trajectories (see Figure 2 and 3) at 4 ℃ (black solid curve), 25 ℃ (red solid curve), 37 ℃ (blue solid curve) and 42 ℃ (pink solid curve) respectively. All curves show a rather linear trend attributed to mostly Brownian motion and a small quadratic component at the beginning (fit to dashed curves in corresponding colors), while at 25 ℃, 37 ℃ and 42 ℃ a second step showing abrupt rise from the linear trend was observed due to convective motion caused by active transport, corresponding to the completion of endocytosis of the single nanotubes. (d) A bar chart showing at different temperatures the change of convective diffusion velocity before (red bars) and after (blue bars) the endocytotic completion point based on the abrupt MSD upturn point. The error bars were obtained by taking the standard deviation of three independent experiments at each temperature.

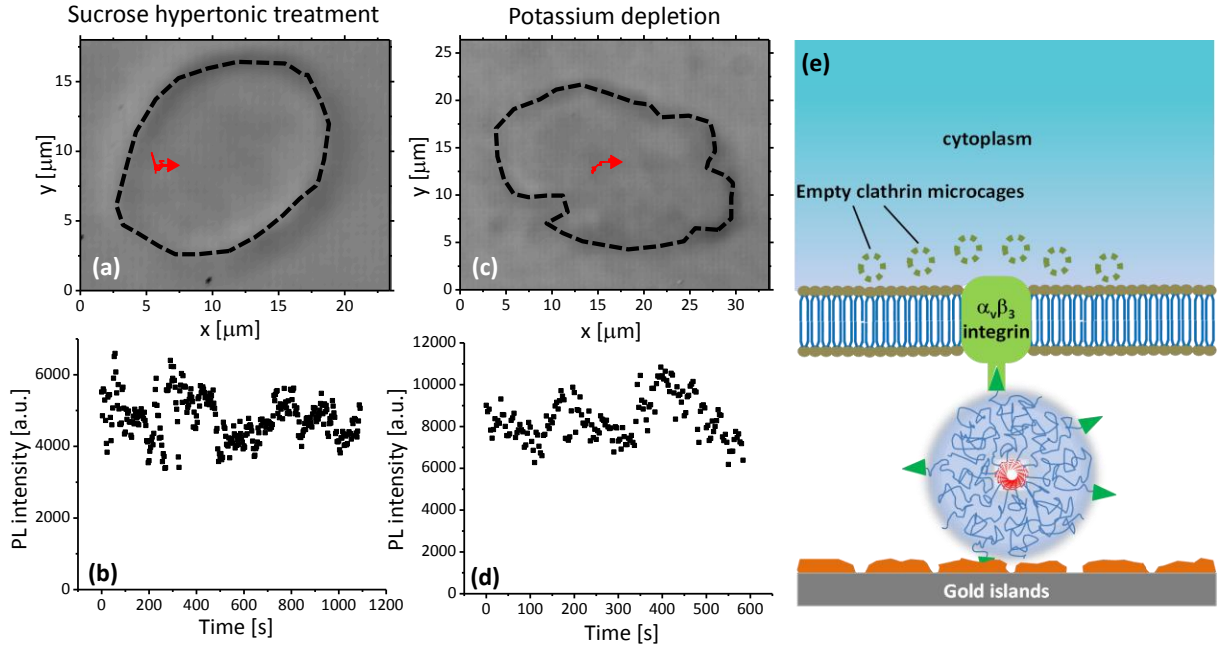

**Figure 5.** Blocking endocytosis of individual nanotubes at 37 ℃. (a)&(c) 2D trajectories of two independent SWCNTs observed on cells pre-treated by sucrose and potassium depletion respectively. (b) & (d) show the time course photoluminescence (PL) of the two individual SWCNTs on the sucrose and potassium depletion treated cells respectively at 37 ℃. (e) A schematic drawing showing the lack of clathrin lattice formation on the cell membrane under either hypertonic or potassium depleted condition, which blocks the formation of clathrin-coated pits (as in Fig.2g,2h) and receptor-mediated endocytosis of the SWCNT.

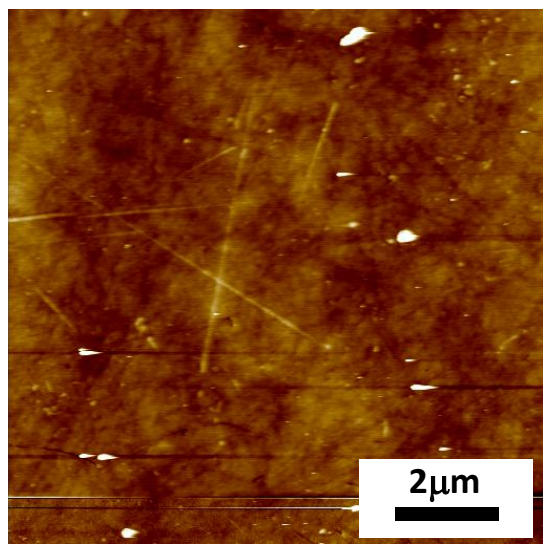

**Supplementary Figure S1.** Length characterization of SWCNTs by AFM. An atomic force microscope (AFM) image of a drop-dried sample on glass, used for calibration curve measurement of the distance dependence of fluorescence enhancement. In this sample, 0.5 uL water-soluble, PEGylated and functionalized SWCNTs (0.45 nM) solution containing 0.05 wt% Triton X-100 was drop-dried on a glass substrate before fluorescence measurement. This sample was calcined at 350 ℃ for 15 min to remove organic matter for AFM.

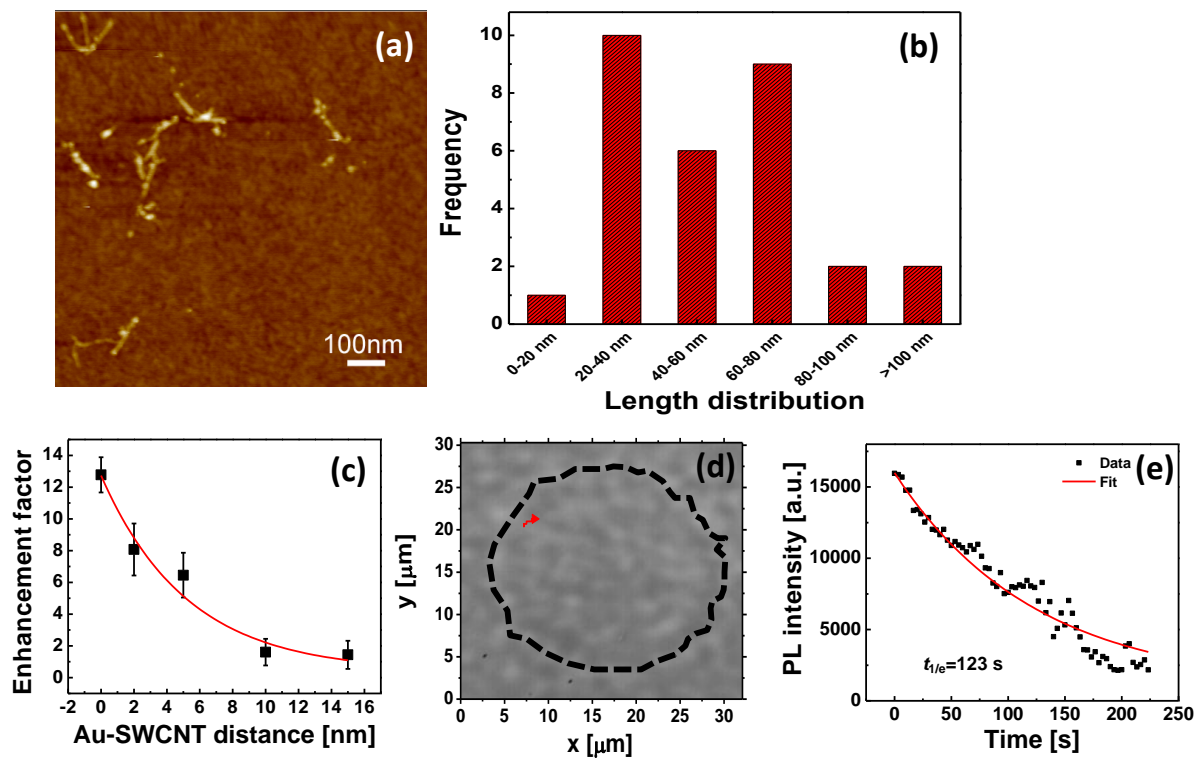

**Supplementary Figure S2.** Fluorescence enhancement and endocytosis imaging of ultrashort SWCNTs on Au/Au film. (a) An AFM image of ultrashort SWCNTs made by sonication and density gradient length separation.[61] (b) A bar chart showing the length distribution of ultrashort SWCNTs with average length of ~53 nm. (c) Fluorescence enhancement factor of ultrashort SWCNTs as a function of distance between SWCNTs and the enhancing Au/Au film. The enhancement is up to ~13-fold and decays with 1/e distance of ~5.5 nm, similar to that of longer tubes (~ 6 nm) used in the main text. Error bars were obtained by taking the standard deviation of all 319×256 pixels in the field of view (690 μm by 860 μm). (d) 2D trajectory of a single ultrashort CNT moving on the U87-MG cell membrane during endocytosis at 37 ℃. (e) Fluorescence intensity of this single ultrashort nanotube plotted as a function of time at 37 ℃, suggesting a similar decay time ($t_{1/e}$ ~ 123 s) for ultrashort nanotube endocytosis compared to longer tubes with $t_{1/e}$ ~ 100~120 s in main text.

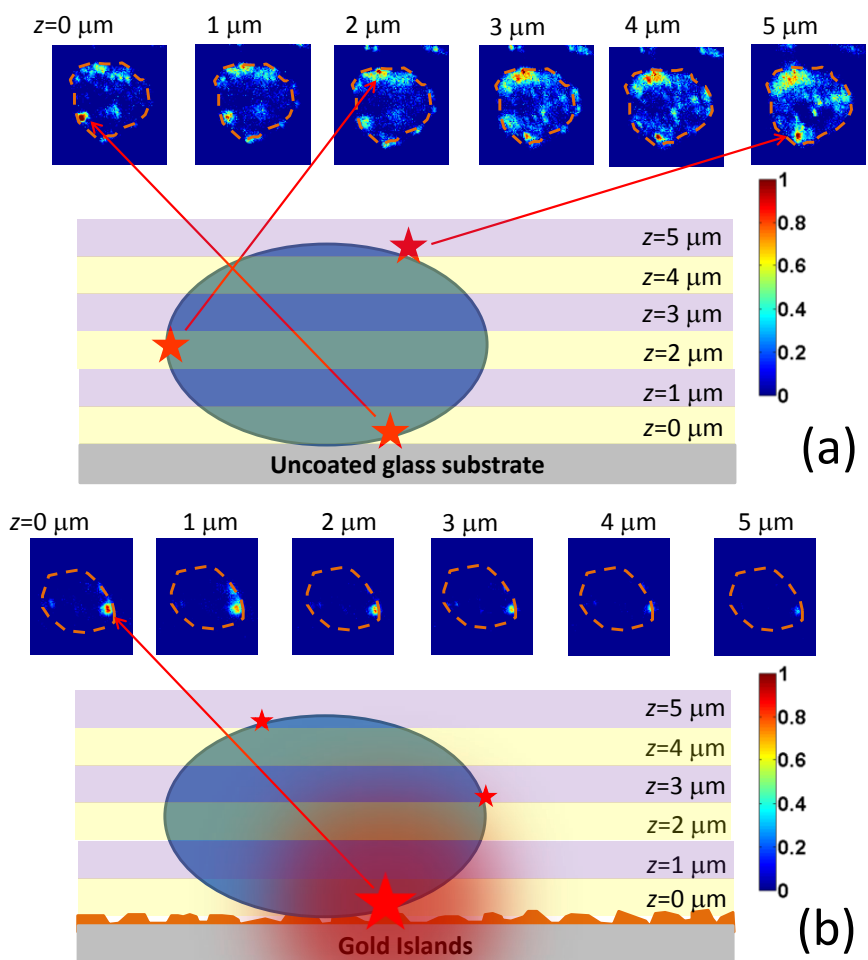

**Supplementary Figure S3.** *Z*-scanned serial images of an SWCNT-stained cell on (a) non-enhancing glass substrate and (b) on Au/Au film (b). The cell on glass was stained at 1 nM of SWCNTs in order to find bright enough tubes while on Au/Au cells could be stained at a much lower 20 pM of SWCNTs for imaging due to the fluorescence enhancement by the Au/Au substrate that helps to visualize the otherwise 'dim' tubes. Note that in (a) different bright spots come into focus at different scanning depths, while in (b) only the bright spot close to gold nano-islands in the Au/Au film is seen and the same spot is seen under the same imaging settings even when focused in other sections since it is the brightest nanotube afforded by highly localized enhancement. Therefore with fixed settings for an SWCNT-stained cell on Au/Au, usually only the nanotubes closer to the enhancing substrate can be seen regardless of the position of the focal plane, which differs from scanning a nanotube stained cell along the *z*-axis on a non-enhancing glass substrate. A scanning step of 1 μm was used to match the depth of field of the Olympus R-4128 ULWD MS Plan 100x IR objective, given by the Rayleigh criteria along the *z*-axis:

$$D_z = \frac{\lambda n}{(NA)^2} = \frac{658 \text{ nm} \times 1.000}{0.80^2} = 1028 \text{ nm}$$

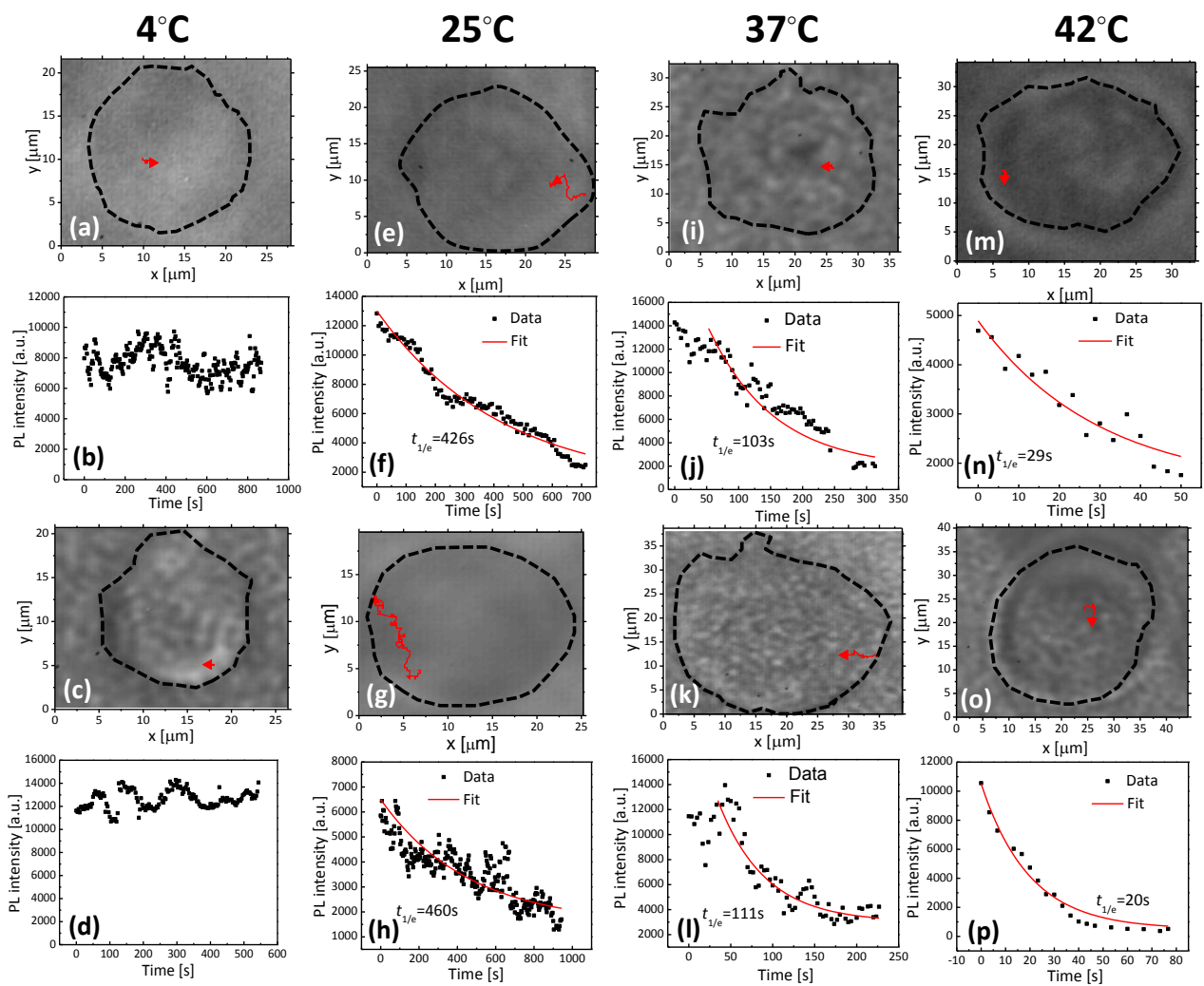

**Supplementary Figure S4.** Reproduced imaging and tracking experiments of individual SWCNTs at 4 ℃ (a-d), 25 ℃ (e-h), 37 ℃ (i-l) and 42 ℃ (m-p) on Au/Au film. 2D trajectories on cell membranes are shown in the first and third row while their corresponding time courses of fluorescence intensity change are shown in the rows underneath.

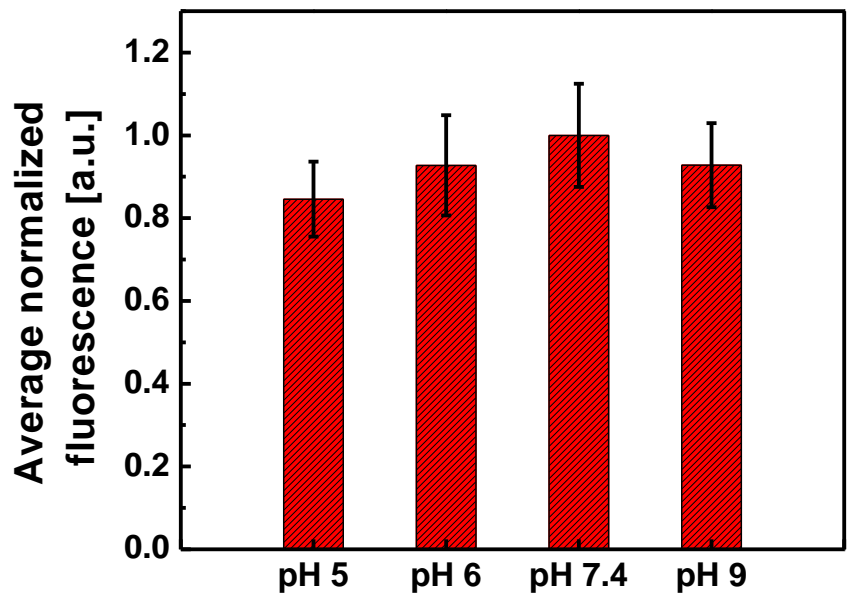

**Supplementary Figure S5.** Average fluorescence intensity of SWCNTs immobilized on Au/Au substrate at different pH at 37 ℃. All intensities were normalized based upon the maximum intensity at pH 7.4. SWCNTs were wrapped in PEG and conjugated with SPDP ligand (N-Succinimidyl 3-[2-pyridyldithio]-propionate) with activatable SH terminus and anchored onto the gold surface via the thiol-gold chemistry. We soaked the Au/Au substrate into SWCNT-PEG-SPDP solution to have enough thiolated SWCNTs adsorbed on the gold surface for ensemble measurement, removed the unbound nanotubes, placed the substrate in 1x PBS at 37 ℃ to mimic the cell imaging condition and imaged the nanotubes through ×100 objective when pH was adjusted between 5 and 9 by adding 0.1 M HCl or 0.1 M NaOH (as the pH of endosomes and lysosomes has been reported as around 5) [42]. Error bars were obtained by taking the standard deviation of all 319×256 pixels in the field of view (69 μm by 86 μm).

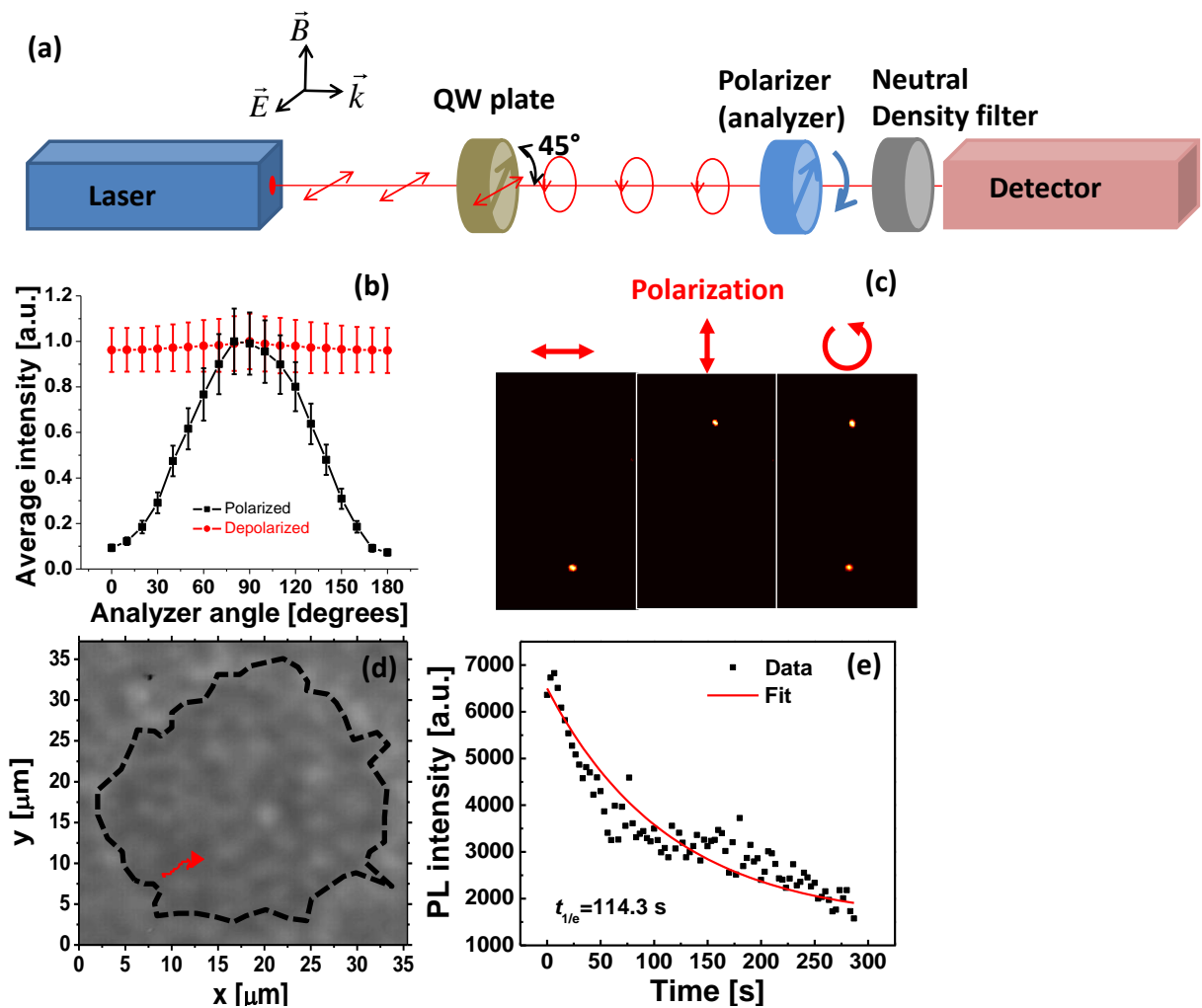

**Supplementary Figure S6.** Depolarization of linearly polarized laser beam for SWCNT excitation. (a) A depolarization setup using a quarter-wave (QW) plate placed at 45 °to the incoming beam polarization. (b) Eccentricity measurements of polarized light without QW plate and depolarized light after QW plate. Error bars were obtained by taking the standard deviation of all 319×256 pixels in the field of view (69 µm by 86 µm). (c) PL images of two perpendicularly oriented single nanotubes upon excitation of linearly and circularly polarized light. (d) 2D trajectory and (e) time course PL intensity change of a single nanotube on cell membranes placed on Au/Au substrate at 37 ℃ imaged using the depolarized light. Similar decay was observed for tubes imaged with linearly polarized light.

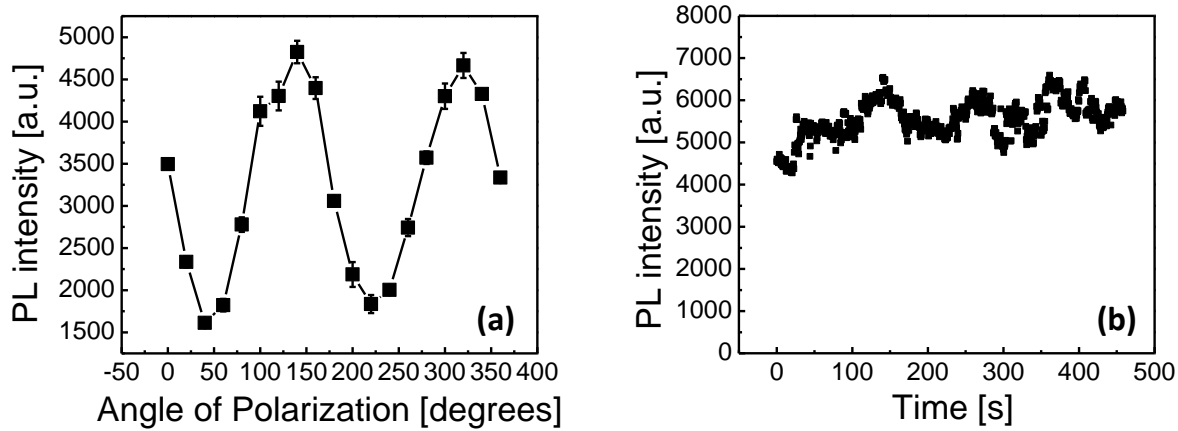

**Supplementary Figure S7.** Fluctuation of photoluminescence (PL) intensity of a single nanotube. (a) Proof of single nanotube by polarization dependent fluorescence emission and (b) Fluorescence intensity over time for a single nanotube strongly anchored on an Au/Au film through thiol-Au linkage (by thiolating SWCNTs using the SPDP crosslinker (N-Succinimidyl 3-[2-pyridyldithio]-propionate), soaking an Au/Au substrate in the nanotube suspension and then immersing the substrate in 1x PBS). Error bars in Supplementary Figure S5a were obtained by taking the standard deviation of the brightest and eight surrounding pixels that make up the single nanotube image.

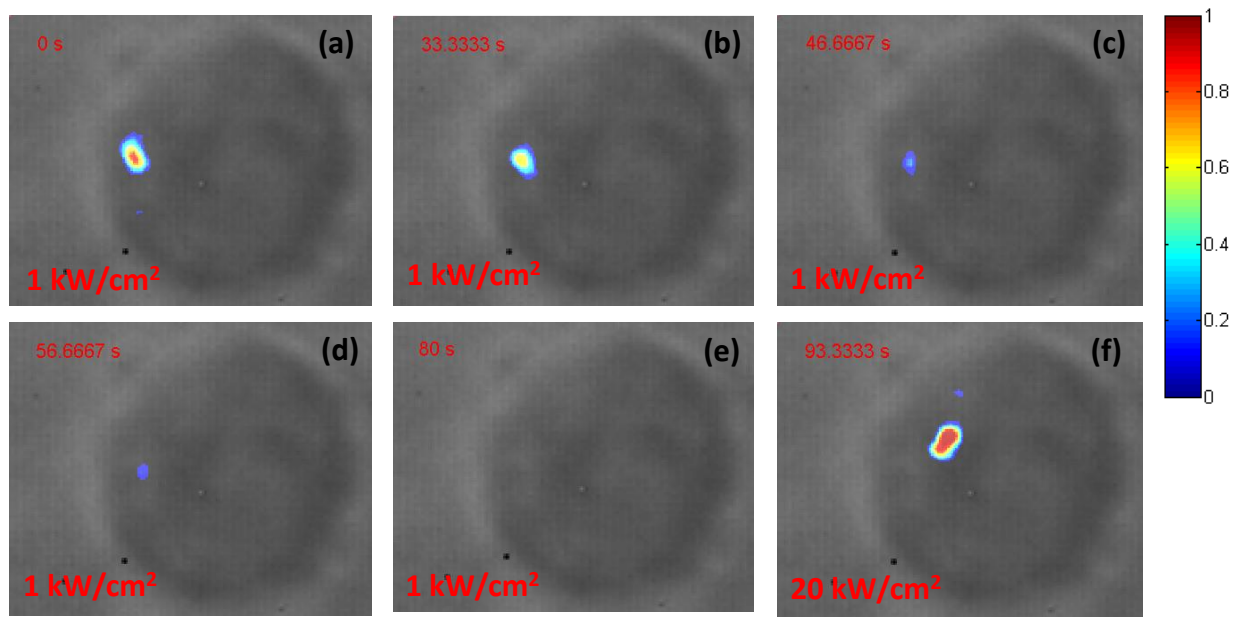

**Supplementary Figure S8.** Tracking a 'disappeared' single nanotube. Fluorescence images of a single nanotube overlaid on optical images of a U87-MG cell at five time points: (a) 0 s, (b) 33.3 s, (c) 46.7 s, (d) 56.7 s, (e) 80 s and (f) 93.3 s after incubation temperature stabilized at 42 ℃. Note that (a-e) were taken under the laser excitation power of 1 kW/cm² while (f) was taken under an increased power of 20 kW/cm² in order to image the 'disappeared' nanotube endocytosed inside the cell away from the plasmonic Au/Au substrate.

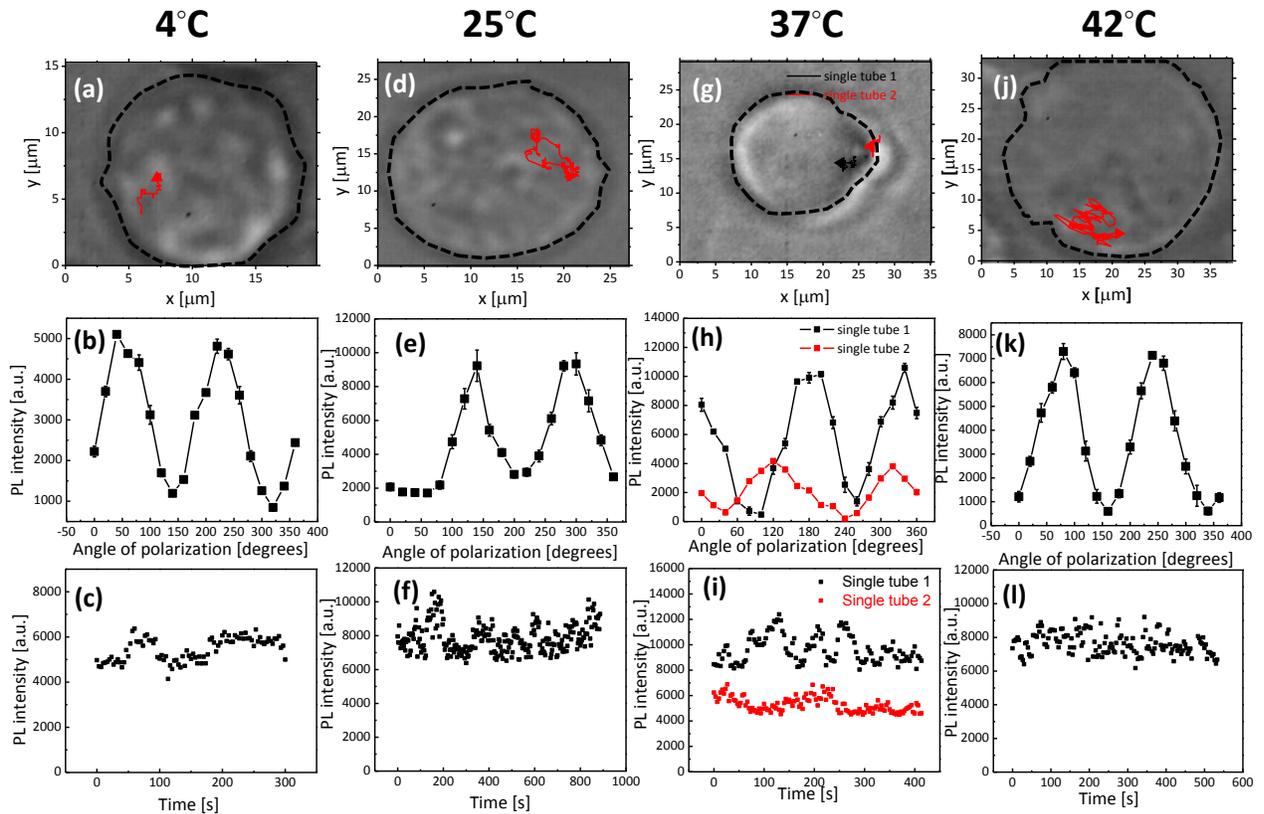

**Supplementary Figure S9.** Single nanotube imaging and tracking at 4 ℃ (a-c), 25 ℃ (d-f), 37 ℃ (g-i) and 42 ℃ (j-l) on glass substrates. 2D trajectories on cell membranes are shown in the first row, proofs of single nanotubes based on polarization dependence are shown in the second row, and their corresponding time courses of fluorescence intensity change are shown in the bottom row. Error bars in all plots of the second row were obtained by taking the standard deviation of the brightest and eight surrounding pixels that make up the single nanotube image.

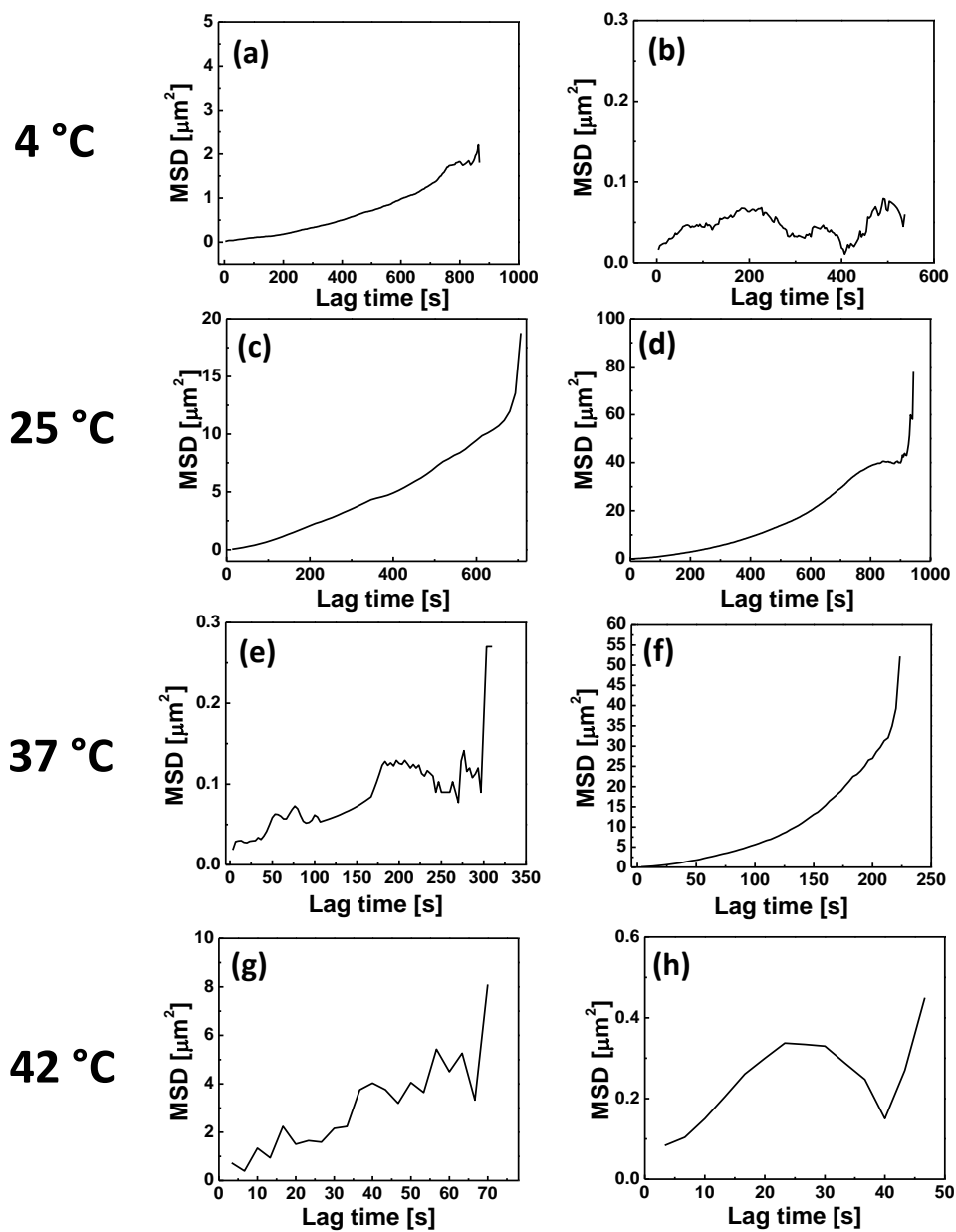

**Supplementary Figure S10.** MSD analysis of other single nanotubes tracked on cell membranes at different temperatures. (a-b) MSD curves vs. lag time from single nanotube tracking at 4 ℃. (c-d) MSD curves vs. lag time from single nanotube tracking at 25 ℃. (e-f) MSD curves vs. lag time from single nanotube tracking at 37 ℃. (g-h) MSD curves vs. lag time from single nanotube tracking at 42 ℃.

# Supplementary References